\documentclass[twocolumn,english,reprint, longbibliography, superscriptaddress, breaklinks=true, showkeys, showpacs=false, nofootinbib]{revtex4}
\usepackage[T1]{fontenc}
\usepackage[latin9]{inputenc}
\usepackage{color}
\usepackage{babel}
\usepackage{amsmath}
\usepackage{amsthm}
\usepackage{amsfonts}
\usepackage{amssymb}
\usepackage{graphicx}
\usepackage{subfigure}
\usepackage{physics}

\usepackage[unicode=true,pdfusetitle,
 bookmarks=true,bookmarksnumbered=false,bookmarksopen=false,
 breaklinks=true,pdfborder={0 0 0},backref=false,colorlinks=true]
 {hyperref} 
\usepackage{breakurl}

\makeatletter
\@ifundefined{textcolor}{}
{%
 \definecolor{BLACK}{gray}{0}
 \definecolor{WHITE}{gray}{1}
 \definecolor{RED}{rgb}{1,0,0}
 \definecolor{GREEN}{rgb}{0,1,0}
 \definecolor{BLUE}{rgb}{0,0,1}
 \definecolor{CYAN}{cmyk}{1,0,0,0}
 \definecolor{MAGENTA}{cmyk}{0,1,0,0}
 \definecolor{YELLOW}{cmyk}{0,0,1,0}
}

\pdfoutput=1
\hypersetup{colorlinks=true,citecolor=blue,linkcolor=cyan,urlcolor=blue,filecolor= green, breaklinks=true}
\usepackage{url}
\usepackage{breakurl}

\makeatother

\begin{document}

\title{Complete complementarity relations in curved spacetimes}

\author{Marcos L. W. Basso}
\email{marcoslwbasso@mail.ufsm.br}
\address{Departamento de F\'isica, Centro de Ci\^encias Naturais e Exatas, Universidade Federal de Santa Maria, Avenida Roraima 1000, Santa Maria, RS, 97105-900, Brazil}

\author{Jonas Maziero}
\email{jonas.maziero@ufsm.br}
\address{Departamento de F\'isica, Centro de Ci\^encias Naturais e Exatas, Universidade Federal de Santa Maria, Avenida Roraima 1000, Santa Maria, RS, 97105-900, Brazil}

\selectlanguage{english}%

\begin{abstract} 
We extend complete complementarity relations to curved spacetimes by considering a succession of infinitesimal local Lorentz transformations, which implies that complementarity remains valid as the quanton travels through its world line and the complementarity aspects in different points of spacetime are connected. This result allows the study of these different complementary aspects of a quantum system as it travels through spacetime. In particular, we investigate the behavior of these different complementary properties of massive spin-$1/2$ particles in the Schwarzschild spacetime. For geodetic circular orbits, we find that the spin state of one particle oscillates between a separable and an entangled state. For non-geodetic circular orbits, we notice that the frequency of these oscillations gets bigger as the orbit gets near to the Schwarzschild radius $r_s$.
\end{abstract}


\maketitle

\section{Introduction}
According to Schr\"odinger, entanglement is the characteristic feature of quantum mechanics, the one that imposes its total departure from the classical lines of thought \cite{Schrodinger}. Its central importance in quantum foundations \cite{Rosen, Bell}, as well as its important role in the fields of quantum information and quantum computation \cite{Popescu, Preskill}, has made entanglement theory achieve great progress in recent decades. Perhaps, the most astonishing application of this unique feature is quantum teleportation, where two observers use two quantum systems in an entangled state to transmit information about the state of a third system \cite{Bennett}. 

Moreover, concern about how entanglement behaves in relativistic scenarios has grown more and more \cite{Terno}. In the end of the last century, Czachor considered the relativistic version of the famous Einstein-Podolsky-Rosen (EPR) experiment with massive spin-1/2 particles \cite{Czachor}. While, in the beginning of this century, the authors of Refs. \cite{Adami, Ueda} showed that the entanglement of Bell states depends on the velocity of the observer. On the other hand, the authors in Ref. \cite{Milburn} argued that the entanglement fidelity of a Bell state remains invariant for a Lorentz boosted observer. However, in the same year, it was demonstrated by Peres \textit{et al.} \cite{Peres} that the entropy of a single massive spin-1/2 particle does not remain invariant under Lorentz boosts. These apparently conflicting results involve systems containing different particle states and boost geometries \cite{Palge}. Therefore, entanglement under Lorentz boosts is highly dependent on the boost scenario in question \cite{Dunningham}, which led to a rich variety of works  by several researchers exploring these different scenarios \cite{Li, Moon, Lee, Jordan, Vedral, Friis, Nasr, Blasone}. More generally, the entanglement for observers constantly accelerated in a flat space-time was considered in Refs. \cite{Alsing, Schuller, Fuentes}. A step forward in the investigations of these relativistic scenarios was taken by Terashima and Ueda \cite{Terashima}, who studied EPR correlations and the violation of Bell's inequalities in curved spacetimes. In addition, the same authors, in Ref. \cite{Ueda1}, studied the decoherence of spin states due to the presence of a gravitational field, by considering a succession of infinitesimal Lorentz transformations. It turns out that decoherence is quite general for a particle in a gravitational field \cite{Kok, Dai}.

However, entanglement is not the only quantum feature that occupies a central position in the world of quantum weirdness. The other feature, known as wave-particle duality, also turns apart the quantum world from the classical world. It is usually considered the main example of Bohr's complementarity principle, which states that quantum systems, or quantons \cite{Leblond}, may possess properties that are equally real but mutually exclusive \cite{Bohr}. Attempts have been made to formalize the wave-particle duality in a quantitative way \cite{Wootters, Engle, Yasin}. In these efforts, quantitative measures of wave and particle properties were constructed and constrained in a complementarity inequality \begin{equation}
    P^2 + V^2 \le 1, \label{eq:cr1}
\end{equation}
where $P$ is the predictability and $V$ is the visibility of the interference pattern. Together with the quantitative formulation of the wave-particle duality, it was noticed that not only extreme cases of full wave and particle natures existing in mutual exclusion is possible, but also intermediate cases of partial wave and particle natures coexisting in a compatibility relation. Until now, many approaches were taken for quantifying the wave-particle properties of a quantum system \cite{Angelo, Coles, Hillery, Qureshi, Maziero}. And, with the development of the field of quantum information, it was suggested that the quantum coherence \cite{Baumgratz} would be a good generalization of the visibility measure \cite{Bera, Bagan,  Mishra}. However, as pointed out by Qian \textit{et al.} \cite{Qian}, complementarity relations like Eq. (\ref{eq:cr1}) do not really capture a balanced exchange between $P$ and $V$ because the inequality permits, for instance, that $V$ decreases due to the interaction of the system with its environment while $P$ can remain unchanged, or even worse, it can decrease together with the visibility of system. It even allows the extreme case $P = V = 0$. Thus, something must be missing from Eq. (\ref{eq:cr1}). As noticed by Jakob and Bergou \cite{Janos}, this lack of knowledge about the system is due to entanglement. This means that the information is being shared with another system and this kind of quantum correlation can be seen as responsible for the loss of purity of each subsystem such that, for pure maximally entangled states, it is not possible to obtain information about the local properties of the subsystems, since we can always purify our system and think of it as part of a multipartite pure quantum system. 

Even though entanglement entropy does not remain invariant under Lorentz boosts, and neither do the measures of predictability and coherence, in Ref. \cite{Marcos} we showed that these three measures taken together, in what is known as a complete complementarity relation (CCR), are Lorentz invariant. Hence, in this work, we extend this result to curved spacetimes by considering a succession of infinitesimal Lorentz transformations, allowing us to study the different complementary aspects of a quanton as it moves through spacetime. It is interesting to see that the situation for complementarity in different reference frames, as well as in different points of a curved spacetime, is similar to the definition of spin states in different Lorentz frames in following sense: Peres \textit{et al.}, in Ref. \cite{Peres}, realized that even though it is possible to formally define spin in any Lorentz frame, there is no direct quantitative relationship between the observable expectation values in different Lorentz frames, i.e., spin is a Lorentz frame-dependent concept. For complementarity, even though it is possible to formally define complementarity in any Lorentz frame or in different points of a curved spacetime, in principle there is no relationship between the complementarity relations of different Lorentz frames or in different points of curved spacetimes. However, our result shows that it is possible to connect complete complementarity relations in different Lorentz frames and in different points in a curved spacetime, which implies that complementarity remains valid as the quanton travels through its world line and the complementarity aspects in different points of spacetime are connected. In addition, we study the behavior of these different complementary aspects of massive spin-$1/2$ particles (or qubits) in the Schwarzschild spacetime. For geodetic circular orbits, we find that the spin-state of one particle oscillates between a separable and an entangled state. For non-geodetic circular orbits, we report that the frequency of these oscillations gets bigger as the orbit gets nearer the Schwarzschild radius, which agrees with the fact that the spin precession near $r_s$ is very rapid, as reported in Ref. \cite{Terashima}. This effect is due to the choice of the tetrad field, and thus to the particular static observer, as will be further discussed in the conclusions (Sec. \ref{sec:con}). 

The organization of this article is as follows. In Sec. \ref{sec:spin}, we discuss the spin dynamics in curved spacetimes by focusing in spin-$1/2$ massive particles. In Sec. \ref{sec:lorccr}, we extend complete complementarity relations for curved spacetimes. Thereafter, in Sec. \ref{sec:relset}, we turn to the study of the behavior of CCR in the Schwarzschild spacetime, by exploring two types of circular orbits. Lastly, in Sec. \ref{sec:con}, we give our conclusions.

\section{Spin Dynamics in Curved Spacetimes}
\label{sec:spin}
\subsection{Spin States in Local Frames }
To study the dynamics of spin-$1/2$ particles in gravitational fields, the use of local frames of reference, which can be defined at each point of spacetime, is required. These local frames are defined through a tetrad field (or vielbein), which is a set of four linearly independent four-vector fields \cite{Wald}. In General Relativity, the gravitational field is encoded in the metric components of a curved spacetime, which is a differential manifold $\mathcal{M}$ \cite{Carroll}. A manifold is simultaneously a very flexible and powerful structure, since it comes equipped naturally with a tangent (or contravariant) and a cotangent (or covariant) vector spaces in each point $p \in \mathcal{M}$, denoted by $T_{p}(\mathcal{M})$ and $T^*_{p}(\mathcal{M})$, respectively. Then, tensor fields of arbitrary rank can be constructed from elements of  $T_{p}(\mathcal{M})$ and $T^*_{p}(\mathcal{M})$ using the tensor product $\otimes$. The differential structure of $\mathcal{M}$ provides, in each point $p$, a coordinate basis for the vector spaces $T_{p}(\mathcal{M})$ and $T^*_{p}(\mathcal{M})$ given by $\{\partial_{\mu}\}$ and $\{ dx^{\nu} \}$, respectively, such that $dx^{\nu}(\partial_{\mu}) := \partial_{\mu}x^{\nu} = \delta^{\ \nu}_{\mu}$. We can proceed by defining a metric $g$ in $\mathcal{M}$, which gives us a Riemannian (or pseudo-Riemaniann) manifold. 

The metric is a covariant tensor field of rank 2, which defines, in each point $p \in \mathcal{M}$, an inner product in $T_p(\mathcal{M)}$ that, in turn, allows us to compute lengths, volumes, angles, time intervals, and so on. Given the basis in $T^*_{p}(\mathcal{M})$, we can express the metric as $g = g_{\mu \nu}(x) dx^{\mu} \otimes dx^{\nu}$, and the elements of the metric which encode the gravitational field are given by $g_{\mu \nu}(x) = g(\partial_\mu, \partial_{\nu})$ \cite{Nakahara}. However, the natural basis $\{\partial_{\mu}\} \subset T_p(\mathcal{M})$ and $\{ dx^{\nu} \} \subset T^*_p(\mathcal{M})$ are not necessarily orthonormal. But we can set up any basis as we like. In particular, we can form an orthonormal basis with respect to the pseudo-Riemannian manifold (spacetime) on which we are working. Following Ref. \cite{Nakahara}, let us consider the linear combination
\begin{align}
    & e_a = e_a^{\ \mu}(x) \partial_\mu, \ \ \ e^a = e^a_{\ \mu}(x)dx^{\mu}, \\
    & \partial_{\mu} = e^a_{\ \mu}(x)e_a, \ \ \ dx^{\mu} = e_a^{\ \mu}(x)e^a.
\end{align}  
To define a local frame at each point $p \in \mathcal{M}$, we require $\{e_a\}$ to be orthonormal in the following sense
\begin{align}
    g(e_a, e_b) := \eta_{ab}, \ \ \ g := \eta_{ab}e^a \otimes e^b,
\end{align}
where $\eta_{ab} = diag(-1,1,1,1)$ is the Minkowski metric. Equivalently, we can define the tetrad field in terms of its components
\begin{align}
    & g_{\mu \nu}(x)e_a^{\ \mu}(x)e_b^{\ \nu}(x) = \eta_{ab},\\ &  \eta_{ab}e^a_{\ \mu}(x)e^b_{\ \nu}(x) = g_{\mu \nu}(x) \label{eq:metr},
\end{align}
with
\begin{align}
    e^a_{\ \mu}(x)e_b^{\ \mu}(x) = \delta^{a}_{\ b}, \ \ \ e^a_{\ \mu}(x)e_a^{\ \nu}(x) = \delta_{\mu}^{\ \nu}.
\end{align}
Here, and from now on, we assumed that Latin letters $a, b, c, d,\cdots$ refers
to coordinates in the local frame; Greek indices $\mu, \nu, \cdots$ runs over the four general-coordinate labels; and repeated indices are to be summed over. 

Furthermore, for general coordinate indices, the lowering and raising of indices is done with the metric $g_{\mu \nu}(x)$ and its inverse $g^{\mu \nu}(x)$, respectively. The indices in the local frame are lowered by $\eta_{ab}$ and raised by its inverse $\eta^{ab}$. The components of the tetrad field and its inverse transforms a tensor in the general coordinate system into one in the local frame, and vice versa. Therefore it can be used to shift the dependence of spacetime curvature of the vector fields to the tetrad fields. Indeed, instead of working with $A^{\mu}$ defined in the general coordinate system, it is possible to work
with $e_a^{\ \mu}(x)A^a$. As $A^a$ is a set of four Lorentz scalar fields, then all the information about the spacetime curvature is encoded in the tetrad field $e_a^{\ \mu}(x)$ \cite{Lanzagorta}. In addition, Eq.(\ref{eq:metr}) tells us that the tetrad field encodes all the information about the spacetime curvature hidden in the metric, which allowed an equivalent formulation of General Relativity in terms of the tetrad fields \cite{Maluf}. Besides, it is worth pointing out that  the tetrad field $\{e_a^{\ \mu}(x), a = 0,1,2,3\}$ is a set of four contravariant vector fields, and not a single second-rank tensor of indices $a$ and $\mu$. Therefore, the tetrad field transforms as
\begin{equation}
    e_a^{\ \mu}(x) \to e_a^{' \ \mu}(x') = \frac{\partial x'^{\mu}}{\partial x^\nu} e_a^{\ \nu}(x)
\end{equation}
under general coordinate transformation, and as
\begin{equation}
    e_a^{\ \mu}(x) \to e_a^{' \ \mu}(x') = \Lambda_a^{\ b}(x) e_b^{\ \mu}(x)
\end{equation}
in the local system, which is a local Lorentz transformation. Since the local frame remains local under local Lorentz transformations, the choice of the local frame is not unique. Therefore, a tetrad representation of a particular metric is not uniquely defined, and different tetrad fields will provide the same metric tensor, as long as they are related by local Lorentz transformations \cite{Misner}.

By using the set of orthonormal four-vectors $e_a^{\ \mu}(x)$, the observer succeed in making the metric components of his laboratory locally flat, $g(e_a, e_b) = \eta_{ab}$. The observer can go even further, constructing coordinates in his laboratory such that the derivative of the metric components $g_{\mu \nu}(x)$ vanishes along the geodetic trajectory of its world-line. Coordinates constructed in this way are known as Riemann normal coordinates, which provide a realization of the locally inertial frames (or freely falling frames). A way to accomplish this is by the exponential map \cite{Carroll}. By constructing the local Lorentz transformation, we can define a particle with spin-$1/2$ in curved spacetimes as a particle whose one-particle states furnish the spin-$1/2$ representation of the local Lorentz transformation \cite{Terashima}. Thus, let us consider a massive spin-$1/2$ particle moving with four-momentum $p^{\mu}(x) = m u^{\mu}(x)$ with $p^{\mu}(x) p_{\mu}(x) = -m^2$, where $m$ is the mass of the quanton, $u^{\mu}(x)$ is the four-velocity in the general coordinate system, and we already put $c = 1$. Now, we can use the tetrad field $e^a_{\ \mu}(x)$ to project the four-momentum $p^{\mu}(x)$ into the local frame, i.e., $p^a(x) = e^a_{\ \mu}(x) p^{\mu}(x)$. Thus, in the local frame at point $p \in \mathcal{M}$ with coordinates $x^a = e^a_{\ \mu}(x) x^{\mu}$, a momentum eigenstate of a Dirac particle in a curved spacetime is given by \cite{Lanzagorta}
\begin{align}
 \ket{p^a(x), \sigma; x} := \ket{p^a(x), \sigma; x^{a}, e^a_{\ \mu}(x), g_{\mu \nu}(x)},
\end{align}
and represents the state with spin $\sigma$ and momentum $p^a(x)$ as observed from the position $x^a = e^a_{\ \mu}(x) x^{\mu}$ of the local frame defined by $ e^a_{\ \mu}(x)$ in the spacetime $\mathcal{M}$ with metric $g_{\mu \nu}(x)$. 

The description of a Dirac particle state can only be provided regarding the tetrad field and the local structure that it describes. By definition, the state $\ket{p^a(x), \sigma; x}$ transforms as the spin-$1/2$ representation under the local Lorentz transformation. In the case of special relativity, a one-particle spin-$1/2$ state $\ket{p^a, \sigma}$ transforms under a Lorentz transformation $\Lambda^{a}_{b}$ as \cite{Weinberg}
\begin{equation}
    U(\Lambda)\ket{p^a, \sigma} = \sum_{\lambda} D_{\sigma \lambda}(W(\Lambda,p)) \ket{\Lambda p^a, \lambda},
\end{equation}
where $D_{\sigma \lambda}(W(\Lambda,p))$ is a unitary representation of the Wigner's little group, whose elements are Wigner rotations $W^{a}_{b} (\Lambda,p)$ \cite{Eugene}. It is worth pointing out that the subscripts in $D_{\sigma, \lambda} (W(\Lambda, p))$ are just to emphasize that in general $U(\Lambda)$ generates superposition in the spin-states.  We could very well suppress the subscripts and write $U(\Lambda) \ket{p^a, \sigma} = \ket{\Lambda p^a} \otimes D (W(\Lambda, p)) \ket{\sigma}$  \cite{Palge}, as sometimes we will do. In other words, under a Lorentz transformation $\Lambda$, the momenta $p^a$ goes to $\Lambda p^a$, and the spin transforms under the representation $D_{\sigma, \lambda}(\Lambda, p)$ of the Wigner's little group \cite{Onuki}. Meanwhile, in a curved spacetime everything above remains essentially the same, except by the fact that single-particle states now form a local representation of the inhomogeneous Lorentz group at each point $p \in \mathcal{M}$, i.e., 
\begin{equation}
    U(\Lambda(x))\ket{p^a(x), \sigma;x} = \sum_{\lambda} D_{\sigma \lambda}(W(x)) \ket{\Lambda p^a(x), \lambda;x} \label{eq:unit},
\end{equation}
where $W(x) := W(\Lambda(x), p(x))$ is a local Wigner rotation.

\subsection{Spin Dynamics}
Following Terashima and Ueda \cite{Terashima}, let us consider how the spin changes when the quanton moves from one point to another in curved spacetime. In the local frame at point $p$ with coordinates $x^a = e^a_{\ \mu}(x) x^{\mu}$, the momentum of the particle is given by $p^a(x) = e^a_{\ \mu}(x) p^{\mu}(x)$. After an infinitesimal proper time $d \tau$, the quanton moves to a new point with general coordinates $x'^{\mu} = x^{\mu} + u^{\mu} d\tau$. Then, the momentum of the particle in the local frame at the new point becomes $ p^a(x') = p^a(x) + \delta p^a(x)$, where the variation of the momentum in the local frame can be described by the combination of changes due to non-gravitational external forces $\delta p^{\mu}(x)$, and spacetime geometry effects $\delta e^a_{\ \mu}(x)$:
\begin{equation}
    \delta p^a(x) = e^a_{\ \mu}(x) \delta p^{\mu}(x) + \delta e^a_{\ \mu}(x)p^{\mu}(x).
\end{equation}
The variation $\delta p^{\mu}(x)$ in the first term on the right hand side of the last equation is simply given by
\begin{equation}
     \delta p^{\mu}(x) = u^{\nu}(x) \nabla_{\nu} p^{\mu}(x) d\tau = m a^{\mu}(x) d\tau, \label{eq:momen}
\end{equation}
where $\nabla_{\nu}$ is the covariant derivative and $a^{\mu}(x):=u^{\nu}(x) \nabla_{\nu} u^{\mu}(x)$ is the acceleration due to a non-gravitational force. 
Once $p^{\mu}(x)p_{\mu}(x) = -m^2$ and $p^{\mu}(x)a_{\mu}(x) = 0$, Eq.(\ref{eq:momen}) can be rewritten as
\begin{equation}
     \delta p^{\mu}(x) = - \frac{1}{m}(a^{\mu}(x)p_{\nu}(x) - p^{\mu}(x)a_{\nu}(x))p^{\nu}(x) d\tau.
\end{equation}
Meanwhile, the variation of the tetrad field is given by
\begin{align}
    \delta e^a_{\ \mu}(x) & = u^{\nu}(x) \nabla_{\nu}e^a_{\ \mu}(x) d\tau \nonumber \\
    & = - u^{\nu}(x) \omega_{\nu \ b}^{\ a}(x) e^b_{\ \mu}(x)d \tau,
\end{align}
where $\omega_{\nu \ b}^{\ a} := e^{a}_{\ \lambda} \nabla_{\nu} e_{b}^{\ \lambda}$ is the connection 1-form (or spin connection) \cite{Chandra}. Collecting these results and substituting in Eq. (\ref{eq:momen}), we obtain
\begin{equation}
    \delta p^a(x) = \lambda^{a}_{\ b}(x)p^{b}(x) d\tau \label{eq:momvar}, 
\end{equation}
where
\begin{align}
    \lambda^{a}_{\ b}(x) = - \frac{1}{m}(a^{a}(x)p_{b}(x) - p^{a}(x)a_{b}(x)) + \chi^{a}_{\ b} \label{eq:infloc}
\end{align}
with $\chi^{a}_{\ b} :=  - u^{\nu}(x) \omega_{\nu \ b}^{\ a}(x)$. It can be shown that Eqs. (\ref{eq:momvar}) and (\ref{eq:infloc}) constitute an infinitesimal local Lorentz transformation since, as the particle moves in spacetime, the momentum in the local frame will transform under an infinitesimal local Lorentz
transformation $p^{a}(x) = \Lambda^{a}_{\ b}(x) p^b(x)$ where $\Lambda^{a}_{\ b}(x) = \delta^{a}_{\ b} + \lambda^{a}_{\ b}(x)d \tau$ \cite{Lanzagorta}. If the particle moves in a geodesic in spacetime, then $a^{\mu}(x) = 0$ and the infinitesimal Lorentz transformation in the local frame reduces to $\lambda^{a}_{\ b}(x) = -u^{\nu}(x) \omega_{\nu \ b}^{\ a}(x)$. 

Now, given the local Lorentz transformation, we can construct the local Wigner rotation that affects the spin of the particle. In other words, by using a unitary representation of the local Lorentz transformation, the state $\ket{p^a(x), \sigma; x}$ is now described as $U(\Lambda(x)) \ket{p^a(x), \sigma; x}$ in the local frame at the point $x'^{\mu}$, and Eq. (\ref{eq:unit}) expresses how the spin of the quanton rotates locally as the particle moves from $x^{\mu} \to x'^{\mu}$ along its world-line. Therefore, one can see that spacetime tells quantum states how to evolve. For the infinitesimal Lorentz transformation, the infinitesimal Wigner rotation is given by
\begin{equation}
    W^{a}_{\ b}(x) = \delta^{a}_{\ b} + \vartheta^{a}_{\ b} d \tau,
\end{equation}
where $\vartheta^{0}_{\ 0}(x) = \vartheta^{i}_{\ 0}(x) = \vartheta^{0}_{\ i}(x) = 0$ and
\begin{equation}
    \vartheta^{i}_{\ j}(x) = \lambda^{i}_{\ j}(x) + \frac{\lambda^{i}_{\ 0}(x)p_j(x) - \lambda_{j0}(x)p^i(x)}{p^0(x) + m}.
\end{equation}
In \cite{Kilian}, the authors provided an explicitly calculation of these elements, and the two-spinor representation of the infinitesimal Wigner rotation is then given by
\begin{align}
    D(W(x)) & = I_{2 \times 2} + \frac{i}{4} \sum_{i,j,k = 1}^{3} \epsilon_{ijk} \vartheta_{ij}(x) \sigma_k d \tau \nonumber \\
    & = I_{2 \times 2} + \frac{i}{2} \boldsymbol{\vartheta} \cdot \boldsymbol{\sigma} d\tau \label{eq:wigner}
\end{align}
where $I_{2 \times 2}$ is the identity matrix, $\{\sigma_k\}_{k = 1}^3$ are the Pauli matrices, and $\epsilon_{ijk}$ is the Levi-Civita symbol. Moreover, the Wigner rotation for a quanton that moves over a finite proper time interval can be obtained by iterating the expression for the infinitesimal Wigner rotation \cite{Terashima}, and the spin-$1/2$ representation for a finite proper time can be obtained by iterating the Eq. (\ref{eq:wigner}):
\begin{equation}
    D(W(x, \tau)) = \mathcal{T}e^{\frac{i}{2}\int_0^{\tau} \boldsymbol{\vartheta} \cdot \boldsymbol{\sigma} d\tau'},
\end{equation}
where $\mathcal{T}$ is the time-ordering operator \cite{Dai}, since, in general, the Wigner rotation varies at different points along the trajectory.

\section{Complementarity relations in Curved Spacetimes}
\label{sec:lorccr}
In Ref. \cite{Basso}, we developed a general framework to obtain a complete complementarity relation (CCR) for a subsystem that belongs to an arbitrary multipartite pure quantum system, by exploring the purity of the multipartite quantum system. While, in Ref. \cite{Marcos}, we demonstrated that this procedure turns out to be useful to prove that the CCR obtained is invariant under Lorentz boosts. In this section, we extend this result to curved spacetimes by considering a succession of infinitesimal Lorentz transformations, as discussed in the previous section. We will restrict ourselves to discrete momentum states, as in Refs. \cite{Jordan, Friis, Palge}, which corresponds to plane waves solutions of the Dirac equation. Besides, this can be justified once we can consider narrow distributions by composing different plane waves solutions such that the momentum states are centered around different momentum values, what makes possible representing them by orthogonal state vectors, i.e., $\braket{p_a}{p_b} = \delta_{a,b} $. Although narrow momenta are an idealization, it is a system worth studying, since it helps to understand more realistic systems. Also, it is worth pointing out that throughout this article we consider only massive particles of spin-$1/2$. By doing this, we are considering a particular representation of the Wigner little group. Besides, we use the standard spin basis, not
other forms like the helicity basis. Even though the helicity basis is more often considered in theoretical and experimental investigations in high energy physics, both the helicity states and the spin states can constitute a basis for the Hilbert space of one particle. As well, both can be studied through unitary representations of the Lorentz group. However, as analyzed recently, the entanglement properties for helicity differ remarkably from those for spin after we trace out the momentum degree of freedom under Lorentz transformations \cite{He, Shao}. Nevertheless, the result obtained in this section will not depend on the particular choice of representation, given that the representation remains unitary.

So, let us consider $n$ massive quantons with spin-$1/2$ in a pure state described by $\ket{\Psi}_{A_1,...,A_{2n}} \in \mathcal{H}_{1} \otimes ... \otimes \mathcal{H}_{2n}$, with dimension $d = d_{A_1}d_{A_2}...d_{A_{2n}}$, in the local frame defined by the tetrad field in the point $p$ of spacetime represented by the coordinates $x^a = e^a_{\ \mu}(x)x^{\mu}$. For instance, $A_1$, $A_2$ are referred as the momentum and spin of the first quanton, and so on.  By defining a local orthonormal basis for each degree of freedom (DOF) $A_m$, $\{\ket{i_m}_{A_m}\}_{i = 0}^{d_m - 1}$, $m = 1,...,2n$, the state of the multipartite quantum system can be written as \cite{Mark}
\begin{align}
    \rho& = \ket{\Psi}_{A_1,...,A_{2n}}\bra{\Psi}\\ & =  \sum_{\overset{i_1,...,i_{2n}}{j_1,...,j_{2n}}} \rho_{i_1 ... i_{2n},j_1...j_{2n}}\ket{i_1,...,i_{2n}}_{A_1,...,A_{2n}} \bra{j_1,...,j_{2n}}. \nonumber
\end{align}
Without loss of generality, let us consider the state of the DOF $A_1$, which is obtained by tracing over the other subsystems:
\begin{align}
    \rho_{A_1} & = \sum_{i_1,j_1}\rho_{i_1,j_1}^{A_1}\ket{i_1}_{A_1}\bra{j_1} \nonumber \\ & = \sum_{\overset{i_1,j_1}{i_2,...,i_{2n}}}\rho_{i_1 i_2 ... i_{2n}, j_1 i_2 ... i_{2n}}\ket{i_1}_{A_1}\bra{j_1}.
\end{align}
The Hilbert-Schmidt quantum coherence  measure \cite{Jonas} of the state $\rho_{A_1}$ is given by
\begin{align}
     C_{hs}(\rho_{A_1}) & = \sum_{i_1 \neq j_1}\abs{\rho_{i_1,j_1}^{A_1}}^2 \nonumber \\
     & = \sum_{i_1 \neq j_1}\abs{\sum_{i_2,...,i_{2n}}\rho_{i_1 i_2 ... i_{2n}, j_1 i_2 ... i_{2n}}}^2,
\end{align}
while the corresponding predictability measure is expressed by 
\begin{align}
        P_{l}(\rho_{A_1}) & = \sum_{i_1}(\rho_{i_1,i_1}^{A_1})^2 - 1/d_{A_1} \nonumber \\
        & = \sum_{i_1}(\sum_{i_2,...,i_{2n}}\rho_{i_1 i_2 ... i_{2n}, i_1 i_2 ... i_{2n}})^2 - 1/d_{A_1}.
\end{align}
We showed in Ref. \cite{Maziero} that these are bona-fide measures of visibility and predictability, respectively. From these equations, an incomplete complementarity relation, $P_{hs}(\rho_{A_1}) + C_{hs}(\rho_{A_1})  \le (d_{A_1} - 1)/d_{A_1}$,  is obtained by exploring the mixedness of $\rho_{A_1}$, i.e., $1 - \Tr \rho_{A_1}^2 \ge 0$. 

The purity of the multipartite quantum system,
$1 - \Tr \rho^2 = 0$, can be written as
\begin{align}
1 - \Big(\sum_{\overset{(i_1,...,i_{2n})}{\overset{=}{(j_1,...,j_{2n})}}}   + \sum_{\overset{(i_1,...,i_{2n})}{\overset{\neq}{(j_1,...,j_{2n})}}}\Big)\abs{\rho_{i_1 i_2 ... i_{2n}, j_1 j_2 ... j_{2n}}}^2 = 0 \label{eq:pur},    
\end{align}
where
\begin{align}
    \sum_{\overset{(i_1,...,i_{2n})}{\overset{\neq}{(j_1,...,j_{2n})}}} & = \sum_{\overset{i_1 \neq j_1}{\overset{i_2 = j_2}{\overset{\vdots}{i_{2n} = j_{2n}}}}} + \sum_{\overset{i_1 = j_1}{\overset{i_2 \neq j_2}{\overset{\vdots}{i_{2n} = j_{2n}}}}} + ... + \sum_{\overset{i_1 = j_1}{\overset{i_2 = j_2}{\overset{\vdots}{i_{2n} \neq j_{2n}}}}} \nonumber \\ & + \sum_{\overset{i_1 \neq j_1}{\overset{i_2 \neq j_2}{\overset{\vdots}{i_{2n} = j_{2n}}}}} + ... + \sum_{\overset{i_1 \neq j_1}{\overset{i_2 = j_2}{\overset{\vdots}{i_{2n} \neq j_{2n}}}}} + ... + \sum_{\overset{i_1 \neq j_1}{\overset{i_2 \neq j_2}{\overset{\vdots}{i_{2n} \neq j_{2n}}}}}.
\end{align}
The linear entropy of the subsystem $A_1$,
\begin{align}
    S_l(\rho_{A_1}) & = 1-Tr\rho_{A_1}^{2} \\
    &=\sum_{\overset{i_1 \neq j_1}{(i_2,...,i_{2n}) \neq (j_2,...,j_{2n})}}\Big(\abs{\rho_{i_1 i_2 ... i_{2n}, j_1 j_2 ... j_{2n}}}^2 \nonumber \\ & -  \rho_{i_1 i_2 ... i_{2n}, j_1 i_2 ... i_{2n}}\rho_{i_1 j_2 ... j_{2n}, j_1 j_2 ... j_{2n}}^*\Big),
    \label{eq:linent}
\end{align}
measures the quantum correlations of $A_1$ with rest of the system. Identifying the predictability, visibility/coherence, and quantum correlations measures within  Eq. (\ref{eq:pur}), we can write down the following CCR:
\begin{align}
    P_{l}(\rho_{A_1}) + C_{hs}(\rho_{A_1}) + S_l(\rho_{A_1}) = \frac{d_{A_1} - 1}{d_{A_1}}. \label{eq:ccrhs}
\end{align}
The proof of this result can be found in Refs. \cite{Marcos, Basso}. It is worthwhile mentioning the CCR given by Eq. (\ref{eq:ccrhs}) is a natural generalization of the complementarity relation obtained by Jakob and Bergou \cite{Jakob, Bergou} for bipartite pure quantum systems. More generally, $E = \sqrt{2S_l(\rho_{A_1})}$, where $E$ is the generalized concurrence obtained in Ref. \cite{Bhaskara} for multi-particle pure states. 

Now, since the dynamics of the quantum system through spacetime can be described by successive local Lorentz transformations, the multipartite quantum system is described by $\ket{\Psi_{\Lambda}}_{A_1,...,A_{2n}} = U(\Lambda(x)) \ket{\Psi}_{A_1,...,A_{2n}}$ at the point $x'^a = e^a_{\ \mu}(x')x'^{\mu}$, and the density matrix of the multipartite pure quantum system can be written as \cite{Caban, Vianna}
\begin{align}
    \rho_{\Lambda} & = \ket{\Psi_{\Lambda}}_{A_1,...,A_{2n}}\bra{\Psi_{\Lambda}} = U(\Lambda(x)) \rho U^{\dagger}(\Lambda(x)), \label{eq:rho}
\end{align}
implying that $\Tr \rho_{\Lambda}^2 = \Tr \rho^2$, and the whole system remains pure as the quantum system moves along its trajectory in spacetime. As we used the purity of the density matrix to obtain the complete complementarity relation for $A_1$, then, from $1 -  \Tr \rho_{\Lambda}^2 = 0$, the following CCR for $A_1$ remains valid throughout the world line of the multipartite quantum system
\begin{equation}
    P_{l}(\rho^{\Lambda}_{A_1}) + C_{hs}(\rho^{\Lambda}_{A_1}) + S_l(\rho^{\Lambda}_{A_1}) = \frac{d_{A_1} - 1}{d_{A_1}}.
\end{equation}
This proves our claim that this complete complementarity relation can be extended to curved spacetimes, allowing us to quantify the different complementary aspects of the subsystems as they move through spacetime.

\section{Qubits in the Schwarzschild Spacetime}
\label{sec:relset}
In this section, we will study the behavior of the different complementary aspects of a spin-$1/2$ quanton (or a qubit), which is in motion in the Schwarzschild spacetime. Because we are interested in qubits, it is worth pointing out that the motion of spinning particles, either classical or quantum, does not follow geodesics because the spin and curvature couples in a non-trivial manner \cite{Papapetrou}. However, the deviation from geodetic motion is very small, of order $\hbar$, and it can be safely ignored except for the case of supermassive compact objects and/or ultra-relativistic test particles \cite{Lanzagorta, Plyatsko, Silenko}. The Schwarzschild solution was the first exact solution to Einstein's field equation, and it describes the spacetime outside of a static and spherically symmetric body of mass M, which constitutes a vacuum solution. Because of its symmetries, the Schwarzschild metric describes a static and spherically symmetric gravitational field \cite{Hobson}. In the spherical coordinates system $(t, r, \theta, \phi)$, the line element of the Schwarzschild metric is given by
\begin{align}
    ds^2 & = g_{\mu \nu}(x)dx^{\mu} dx^{\nu} \\
    & = -f(r)dt^2 + f^{-1}(r)dr^2 + r^2(d\theta^2 + \sin^2 \theta d\phi^2), \nonumber 
\end{align}
where $f(r) = 1 - r_s/r$, with $r_s = 2GM$ being the Schwarzschild radius. It's straightforward to observe that the metric diverges in two distinct points, at $r = r_s$ and at $r = 0$. However, it is important to distinguish the different nature of both singularities. Since all the information about the physics and the spacetime curvature is contained in the curvature tensor $R^{\alpha}_{\ \beta \mu \nu}$ and its contractions, to establish when a metric has a singularity with some physical meaning, it is necessary to search for non-trivial scalars that can be constructed from the curvature tensor, which are independent from coordinate systems. For instance, $R^{\alpha \beta \mu \nu}R_{\alpha \beta \mu \nu} = 12r^2_s/r^6$ tell us that there exists a singular point in $r = 0$ \cite{Carroll}. This suggests that the singularity at $r = r_s$ is not an
intrinsic singularity, since it can be shown that all curvature scalars are finite at $r = r_s$. This type of singularity is called apparent singularity (or coordinate singularity) and it is related to our specific choice of coordinates. Therefore, it can be removed by changing the coordinate system. 

To make the Schwarzschild metric reduce to the Minkowski metric, it is possible to choose the following tetrad field
\begin{align}
    & e^0_{\ t}(x) = \sqrt{f(r)}, \ \ \ e^1_{\ r}(x) = \frac{1}{\sqrt{f(r)}} \nonumber, \\
    & e^2_{\ \theta}(x) = r, \ \ \ e^3_{\ \phi}(x) = r \sin \theta,
\end{align}
and all the other components are zero. Also, only nonzero components will be shown from now on. The inverse of these elements are given by 
\begin{align}
    & e_0^{\ t}(x) = \frac{1}{\sqrt{f(r)}}, \ \ \ e_1^{\ r}(x) = \sqrt{f(r)} \nonumber, \\
    & e_2^{\ \theta}(x) = \frac{1}{r}, \ \ \ e_3^{\ \phi}(x) = \frac{1}{r \sin \theta}.
\end{align}
Thus, we can write the line element as
\begin{align}
    ds^2 & = g_{\mu \nu}(x)dx^{\mu} dx^{\nu} = g_{\mu \nu}(x) e_a^{\ \mu}(x)e_b^{\ \nu}(x) e^a e^b \nonumber \\
    & =  \eta_{ab}e^a e^b.
\end{align} 
This vierbein represents a static local frame at each point. Therefore it can used to represent an observer in the associated local frame \cite{Terashima}. It is worth pointing out that such static local frame is not inertial. In addition, at each point, the $0-, 1-, 2-,$ and $3-$axes are parallel to the $t, r, \theta,$ and $\phi$ directions, respectively. A straightforward calculation shows that the non-zero components of the connection 1-form $\omega_{\nu \ b}^{\ a} := e^{a}_{\ \lambda} \nabla_{\mu} e_{b}^{\ \lambda}$ are given by
\begin{align}
    & \omega_{t \ 0}^{\ 1}(x) = \omega_{t \ 0}^{\ 0}(x) = \frac{r_s}{2r^2},\\
    & \omega_{\theta \ 2}^{\ 1}(x) = - \omega_{\theta \ 1}^{\ 2}(x) = -\sqrt{f(r)},\\
    & \omega_{\phi \ 3}^{\ 1}(x) = - \omega_{\phi \ 1}^{\ 3}(x) = -\sqrt{f(r)}\sin \theta,\\
    & \omega_{\phi \ 3}^{\ 2}(x) = - \omega_{\phi \ 2}^{\ 3}(x) = -\cos \theta.
\end{align}
Now that we have the ingredients required to study the behavior of the Wigner rotation and the different aspects of a qubit in motion in the Schwarzschild spacetime, in the following subsections we will consider two examples: (1) an equatorial circular geodesic and (2) a non-geodetic equatorial circular orbit.

\subsection{Equatorial Circular Geodesics}
Following Refs. \cite{Lanzagorta, Hobson}, let us consider the case of a free-falling test particle moving around the source of the gravitational field in a geodetic circular orbit, which can be obtained by solving the geodesic equation. The four-velocity of these circular geodesics in the equatorial plane, $\theta = \pi/2$, are given by:
\begin{align}
    & u^t = \frac{K}{f(r)}, \ \ \ u^r = 0, \\
    & u^{\theta} = 0, \ \ \ \ \ u^{\phi} = \frac{J}{r^2}, 
\end{align}
where $K, J$ are integration constants related to the energy and angular momentum of the required orbit, respectively, and are given by
\begin{align}
    K = \frac{1 - r_s/r}{\sqrt{1 - \frac{3r_s}{2r}}}, \ \ \ J^2 = \frac{1}{2} \frac{r r_s}{1 - \frac{3r_s}{2r}}.
\end{align}
The energy of the spin-$1/2$ quanton of rest mass $m$ in a circular orbit of radius $r$ is then given by $E = K m$. Furthermore, the value of $J$ implies that the angular velocity is given by 
\begin{equation}
    u^{\phi} = \pm \sqrt{\frac{r_s}{2r^3(1 - \frac{3r_s}{2r})}},
\end{equation}
which means that stable circular geodesic orbits are only possible when $r > \frac{3}{2}r_s$. The non-zero infinitesimal Lorentz transformations in the local frame defined by the tetrad field are given by
\begin{align}
    & \lambda^{0}_{\ 1} = \lambda^{1}_{\ 0} = - \frac{K r_s}{2r^2f(r)}\\ 
    & \lambda^{1}_{\ 3} = - \lambda^{3}_{\ 1} = \frac{J \sqrt{f(r)}}{r^2},
\end{align}
which corresponds to a boost in the direction of the 1-axes and a rotation over the 2-axis, respectively. While, the four-velocity in the local frame is found to be
\begin{equation}
    u^a = e^a_{\ \mu}(x)u^{\mu} = \Big(\frac{K}{\sqrt{f(r)}}, 0, 0, \frac{J}{r}\Big).
\end{equation}
Therefore, the Wigner angle that corresponds to the rotation over the 2-axis is given by:
\begin{align}
      \vartheta^{1}_{\ 3}(x) & = \lambda^{1}_{\ 3}(x) + \frac{\lambda^{1}_{\ 0}(x)p_3(x) - \lambda_{30}(x)p^1(x)}{p^0(x) + m}\\
      & = \frac{J \sqrt{f(r)}}{r^2} \Big(1 - \frac{K r_s}{2rf(r)}\frac{1}{K + \sqrt{f(r)}}\Big).
\end{align}
After the test particle has moved in the circular orbit across some proper time $\tau$, the total angle is given by
\begin{align}
    \Theta & = \int \vartheta^{1}_{\ 3}(x) d \tau = \int \vartheta^{1}_{\ 3}(x) \frac{d \tau}{d \phi} d\phi \\
    & = \frac{\vartheta^{1}_{\ 3}(x)r^2}{J} \Phi, 
\end{align}
since, for a circular orbit, $r$ is fixed and $\vartheta^{1}_{\ 3}(x), K,$ and $J$ are constants. The angle $\Phi$ is the angle traversed by the particle during the proper time $\tau$. It is noteworthy that the angle $\Theta$ reflects all the rotation suffered by the spin of the qubit as it moves in the circular orbit, which means that there are two contributions: The ``trivial rotation'' $\Phi$ and the rotation due to gravity \cite{Terashima}. Therefore, to obtain the Wigner rotation angle that is produced solely by spacetime effects, it is necessary to compensate the trivial rotation angle $\Phi$, i.e., $\Omega := \Theta - \Phi$ is the total Wigner rotation of the spin exclusively due to the spacetime curvature, which only depends on the radius of the circular geodesic $r$ and the mass of the source of the gravitational field expressed by $r_s$.

Similarly to Terashima and Ueda \cite{Terashima}, let us consider a pair of entangled spin-$1/2$ particles emitted at given a point on a geodesic equatorial circle with the local quantization axis along the 1-axis, as one of the particles of the bipartite state circulates the orbit clockwise, the other circulates it counterclockwise. In other words, we have a pair of entangled particles moving in opposite directions with constant four-velocity $u^a_{\pm} = (K/\sqrt{f(r)}, 0, 0, \pm J/r)$ and in the following initial state
\begin{align}
    \ket{\Psi}_{A,B}  =& \frac{1}{\sqrt{2}}\Big(\ket{p^a_+,\uparrow;0}_A \otimes \ket{p^a_-,\downarrow;0}_B \nonumber \\& + \ket{p^a_-,\downarrow;0}_A \otimes \ket{p^a_+,\uparrow;0}_B\Big) \label{eq:state},
\end{align}
where $\phi = 0$ is the coordinate of the point where the quantons were emitted. After some proper time $\tau = r^2 \Phi/J$, the particles travelled along its circular paths and the spinor representation of the finite Wigner rotation due only to gravitation effects is given by
\begin{align}
    D(W(\pm \Phi)) = e^{\mp \frac{i}{2}\sigma_2 \Omega} \label{eq:wigrot},
\end{align}
since $\vartheta^{1}_{\ 3}(x)$ is constant along the path, the time-ordering operator is not necessary. Therefore, the state of the bipartite system in the local frame at points $\phi = \pm \Phi$ is given 
\begin{align}
\ket{\Psi_\Lambda}_{A,B} =& \frac{1}{\sqrt{2}}\cos \frac{\Omega}{2} \sin \frac{\Omega}{2}\Big(\ket{p^a_{+\Phi}, p^a_{-\Phi}}_{A,B} \nonumber
\\ & + \ket{p^a_{-\Phi}, p^a_{+\Phi}}_{A,B}\Big)\otimes\Big(\ket{\uparrow, \uparrow}_{A,B} + \ket{\downarrow, \downarrow}_{A,B}\Big)  \nonumber \\ & + \frac{1}{\sqrt{2}}\ket{p^a_{+\Phi}, p^a_{-\Phi}}_{A,B} \otimes \Big(\cos^2 \frac{\Omega}{2} \ket{\uparrow, \downarrow}_{A,B} \nonumber \\& + \sin^2 \frac{\Omega}{2} \ket{\downarrow, \uparrow}_{A,B}\Big) + \frac{1}{\sqrt{2}}\ket{p^a_{-\Phi}, p^a_{+\Phi}}_{A,B}\nonumber \\& \otimes \Big(\sin^2 \frac{\Omega}{2} \ket{\uparrow, \downarrow}_{A,B}  + \cos^2 \frac{\Omega}{2} \ket{\downarrow, \uparrow}_{A,B}\Big), \label{eq:state1}
\end{align}
where $\ket{p^a_{+\Phi}, p^a_{-\Phi}}_{A,B} := \ket{p^a_{+}; \Phi}_A \otimes \ket{p^a_{-}; -\Phi}_B$. Whereas the reduced spin density matrices of each particle are given by
\begin{align}
        & \rho^A_{\Lambda s} = \rho^B_{\Lambda s} = \begin{pmatrix}
\frac{1}{2} & \cos \frac{\Omega}{2} \sin \frac{\Omega}{2}  \ \\
\cos \frac{\Omega}{2} \sin \frac{\Omega}{2} & \frac{1}{2}
\end{pmatrix}, \label{eq:cohen}
\end{align}
and $\rho^A_{\Lambda p} = \rho^B_{\Lambda p} = \frac{1}{2} I_{2 \times 2}$. By inspecting Eq. (\ref{eq:cohen}), we can see that part of the entanglement between the spins were turned into quantum coherence of each spin state.

\begin{figure}[t]
    \centering
    \subfigure{{\includegraphics[width=7.7cm]{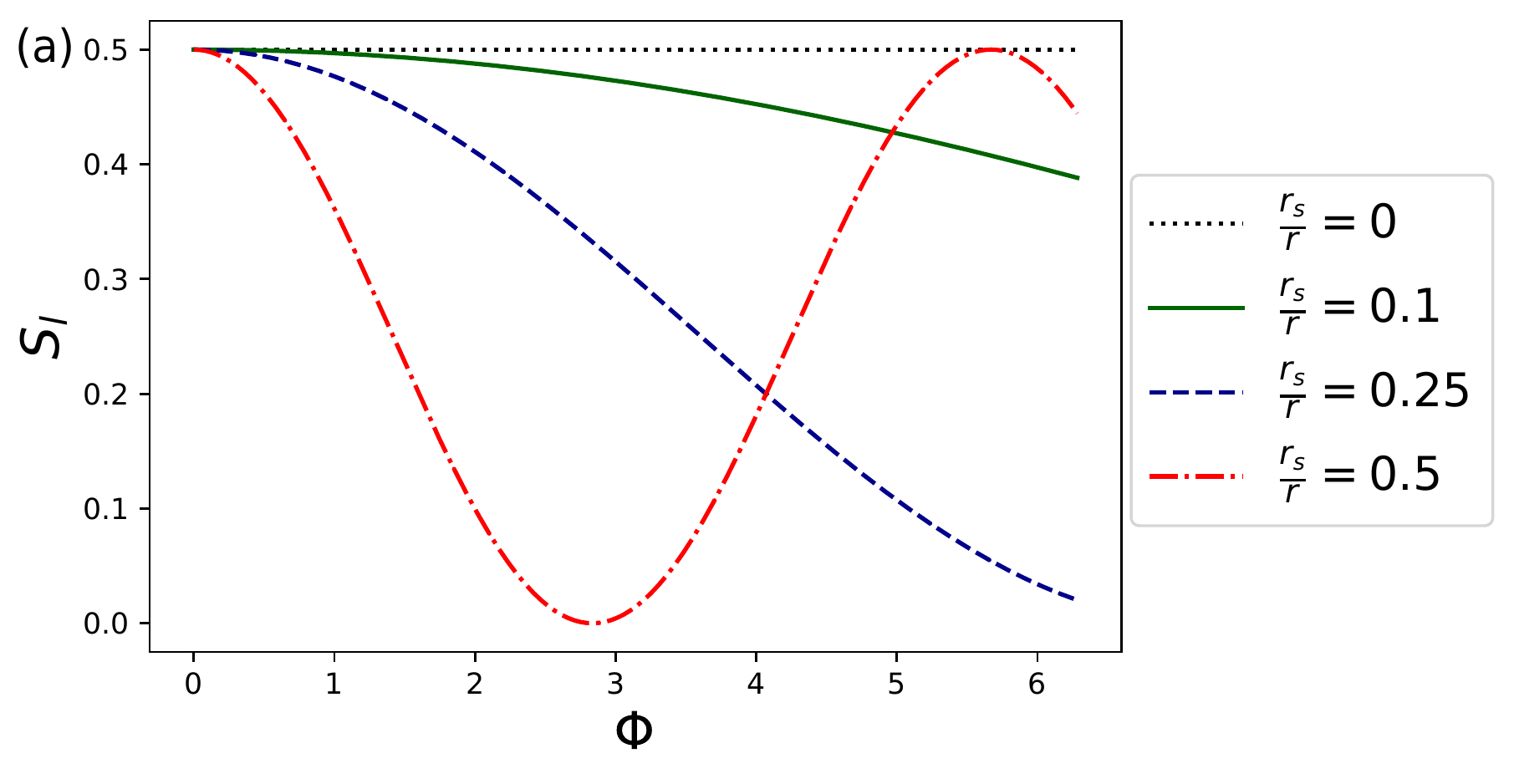}{\label{fig:a}} }}
    \qquad
    \subfigure{{\includegraphics[width=7.7cm]{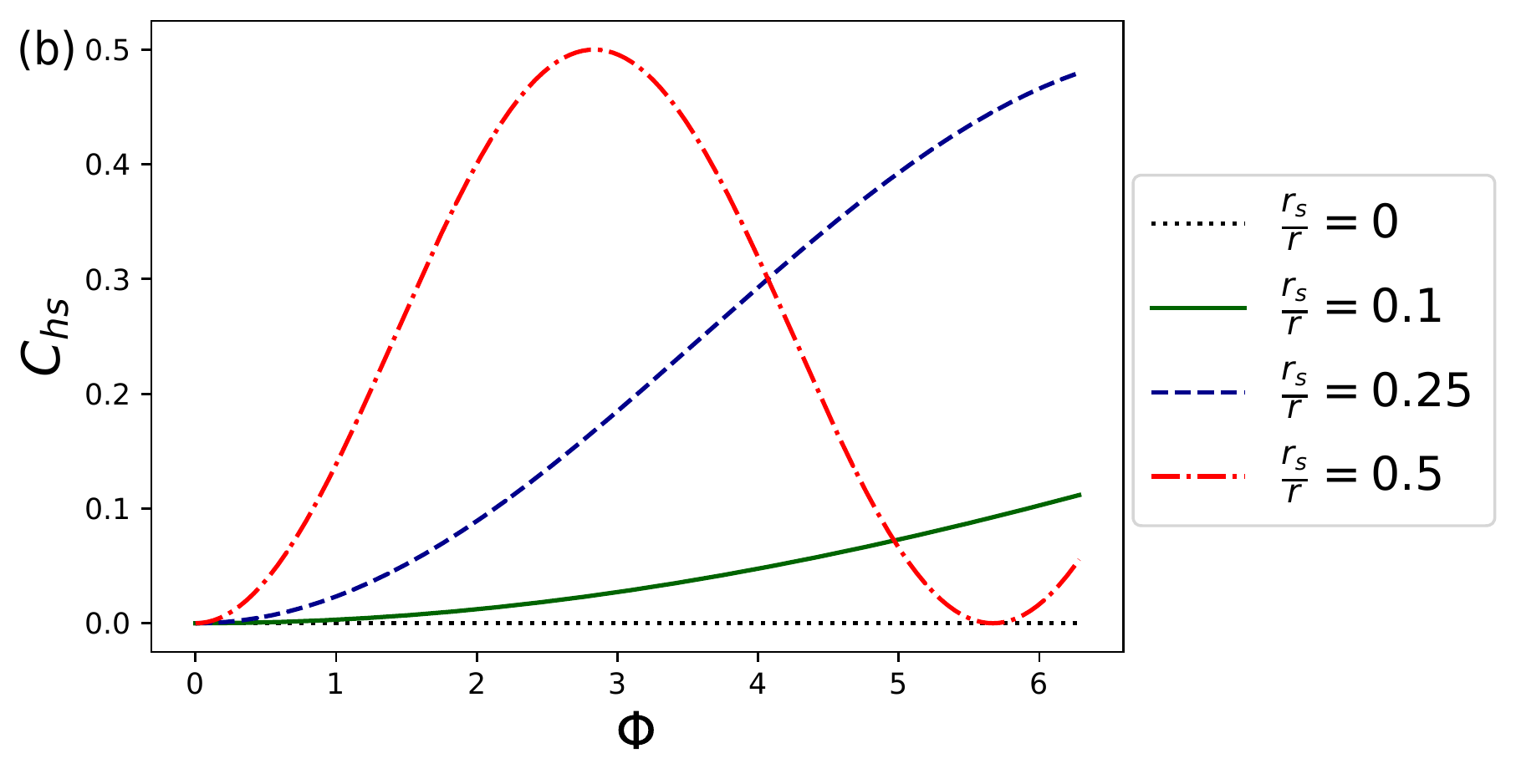}{\label{fig:b}} }}
    \qquad
    \subfigure{{\includegraphics[width=7.7cm]{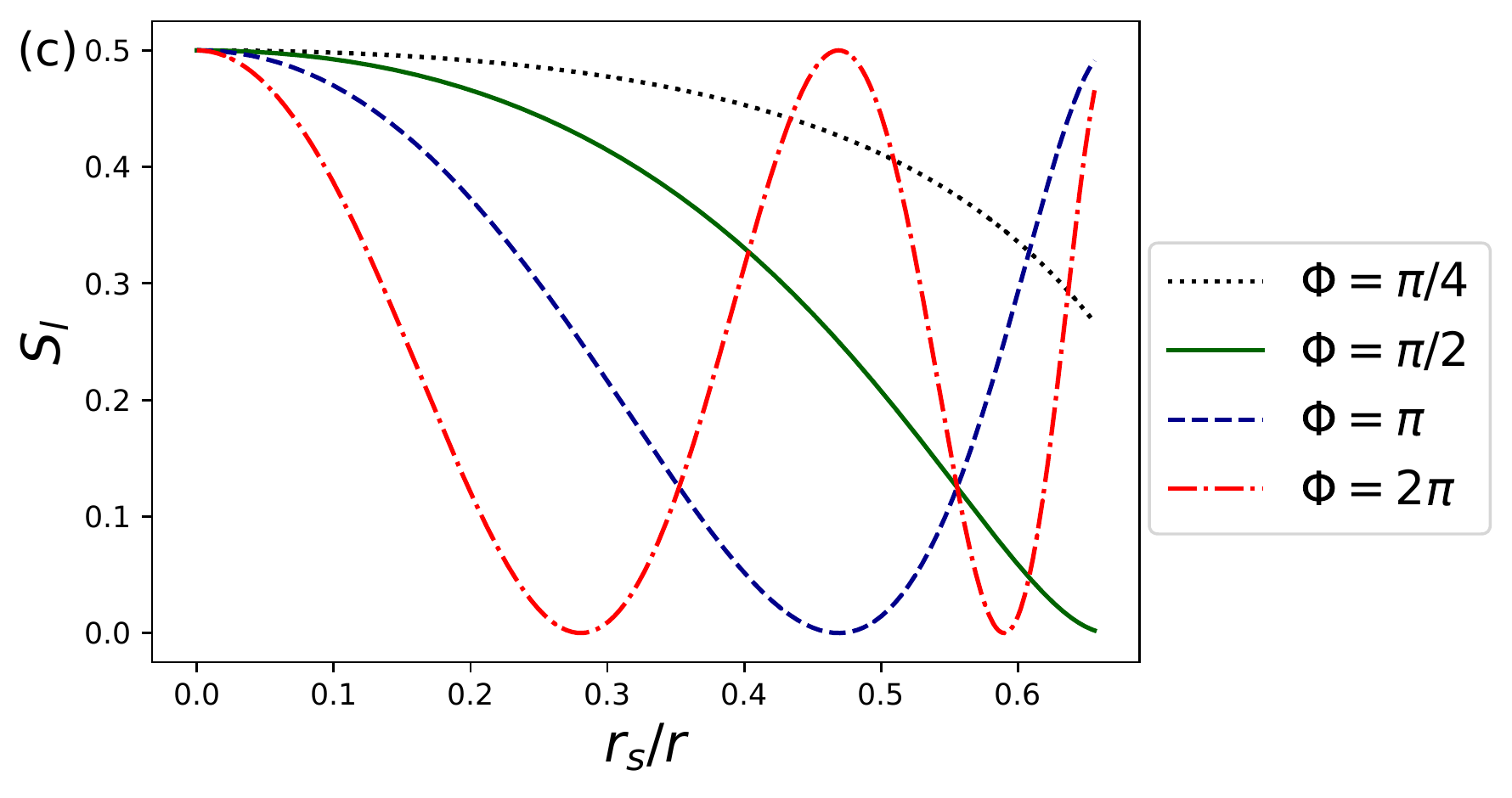}{\label{fig:c}} }}
    \qquad
    \subfigure{{\includegraphics[width=7.7cm]{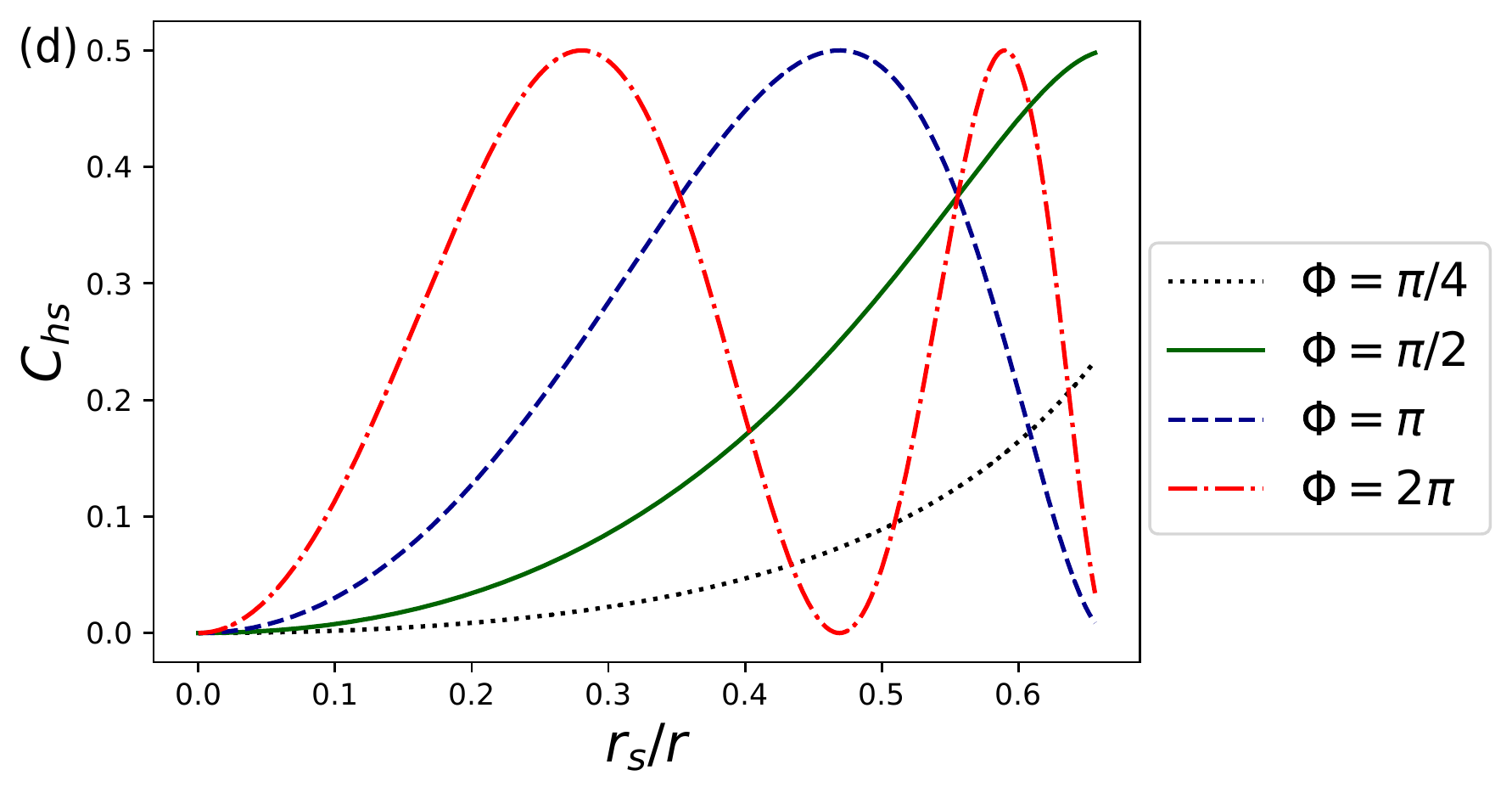}{\label{fig:d}} }}
    \label{fig:example}
    \caption{(Color online) (a) $S_l$, (b) $C_{hs}$ as a function of $\Phi$ and (c) $S_l$, (d) $C_{hs}$ as a function of $r_s/r$.  Quantum coherence and correlation, and their complementarity, for the state in Eq. (\ref{eq:state1}).}
\end{figure}

In Figs. \ref{fig:a} and \ref{fig:b}, we plotted $S_l(\rho^A_{\Lambda s})$ and $C_{hs}(\rho^A_{\Lambda s})$ as a function of $\Phi$ for different circular orbits. As $\Phi \propto \tau$, these figures show the behavior of $S_l(\rho^A_{\Lambda s})$ and $C_{hs}(\rho^A_{\Lambda s})$ as the particle travels along its circular orbit. It is interesting noticing that for $r = 2r_s$, the spin-state of the quanton $A$ oscillates between a separable and an entangled state with the spin-state of the particle $B$, if both particles complete a circular orbit. This behavior is due to the fact that as $r \to \frac{3}{2}r_s$, the Wigner rotation $\Omega$ varies more rapidly. While, in Figs. \ref{fig:c} and \ref{fig:d}, we plotted $S_l(\rho^A_{\Lambda s})$ and $C_{hs}(\rho^A_{\Lambda s})$ as a function of $r_s/r$ for different values of $\Phi$. 

On the other hand, if we consider a one-particle state in a separable state between spin and momentum with a maximally coherent momentum-state in the clock- and counterclockwise directions and the spin-state also maximally coherent, then $C_{hs}(\rho_{\Lambda s})$ would start in its maximum value and decrease, while $S_l(\rho_{\Lambda s})$ would start in its minimum value and increase. Similarly, if we consider a one-particle state in a separable state with a momentum state maximally coherent in the clock- and counterclockwise direction and a spin-state completely predictable, for instance $\ket{\uparrow}$, as the particle travels along its superposition paths, the states of the momenta will become entangled with the spin-states, and there will be a interchange between predictability $P_l(\rho_{\Lambda s})$, and the entanglement entropy $S_l(\rho_{\Lambda s})$ of the spin states. This effect of spacetime curvature in the complementary behavior of these quantum states is analogous to the effect reported in Ref. \cite{Zych}, since each clockwise and counterclockwise circular path can be taken as the different path of a Mach-Zehnder interferometer.

\subsection{Non-geodetic Equatorial Circular Orbit}
\begin{figure}[htp]
    \centering
    \subfigure{{\includegraphics[width=8cm]{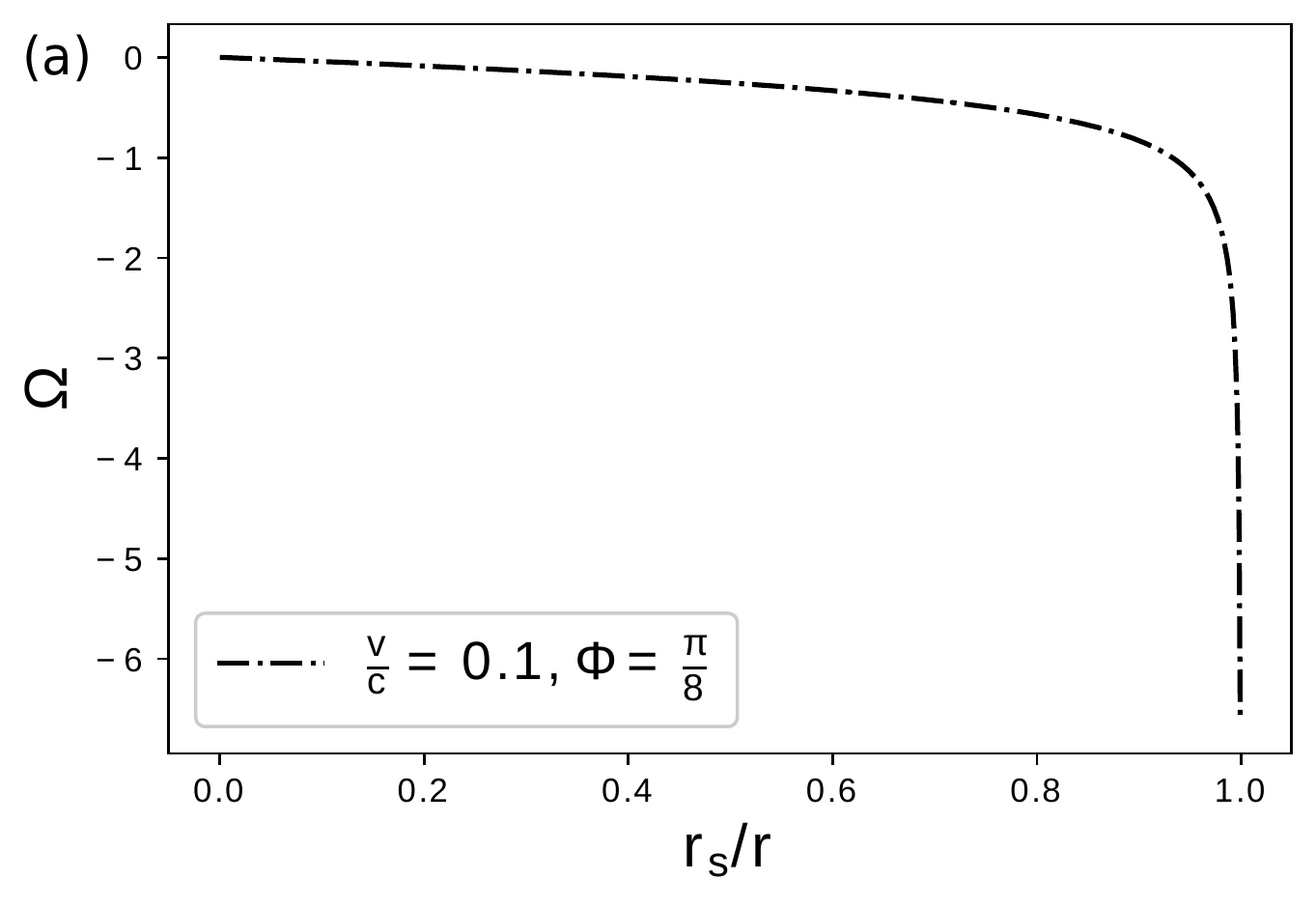}{\label{fig:e}} }}
    \qquad
    \subfigure{{\includegraphics[width=8cm]{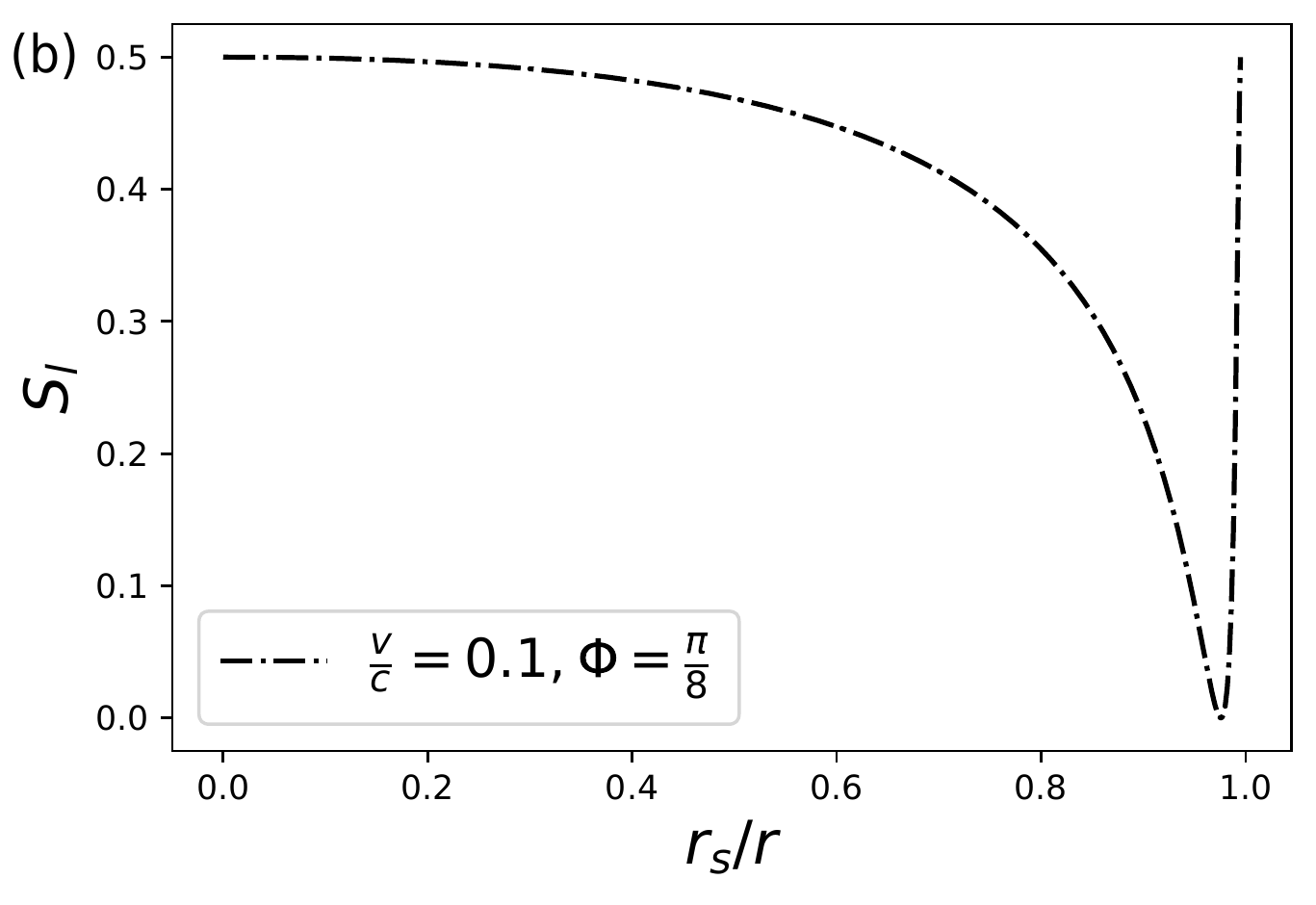}{\label{fig:f}} }}
    \qquad
    \subfigure{{\includegraphics[width=8cm]{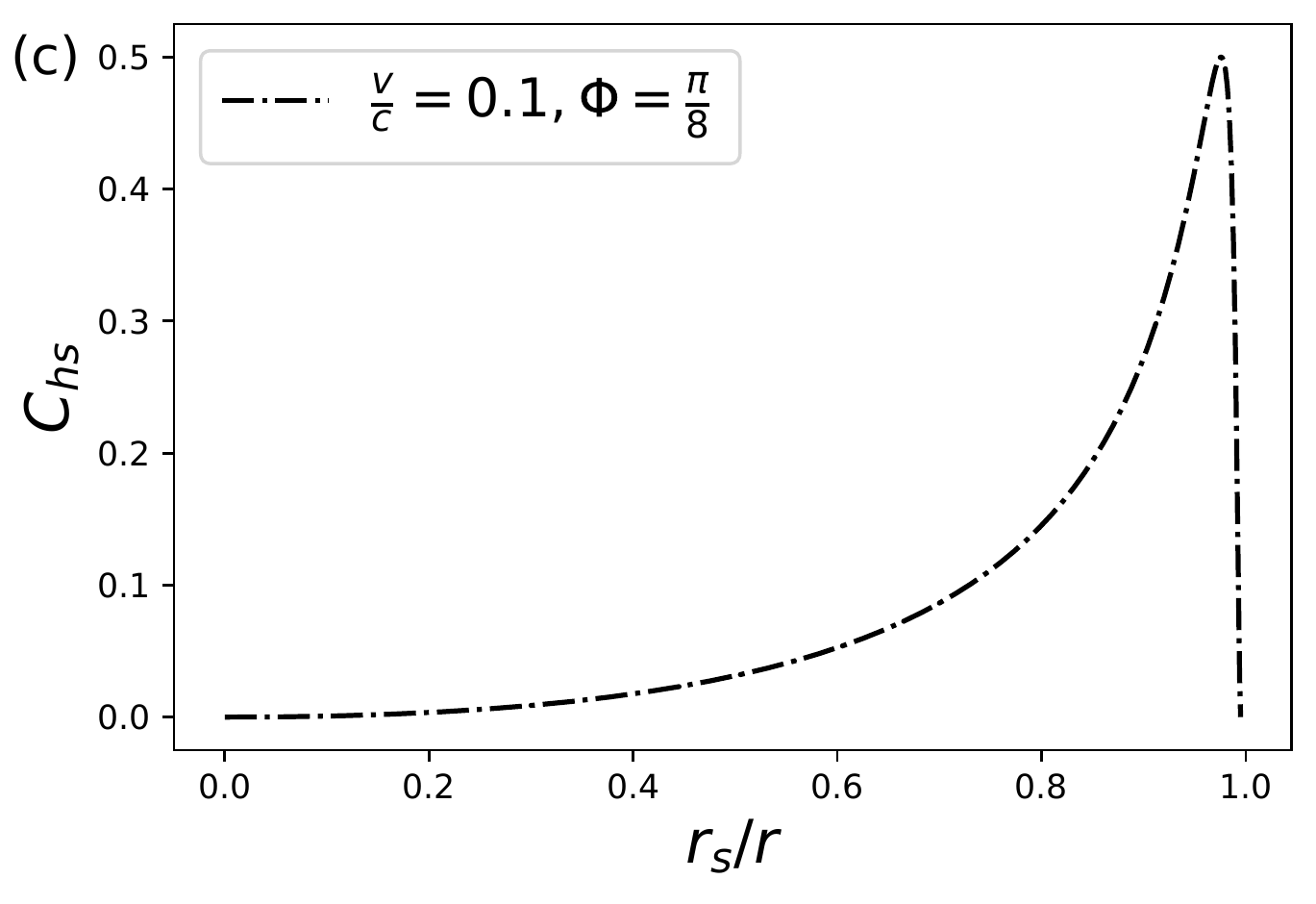}{\label{fig:g}} }}
    \caption{(a) $\Omega$, (b) $S_l$ and (c) $C_{hs}$ as a function of $r_s/r$ The angle $\Omega$, quantum coherence, and quantum correlation, and their complementarity, for the state in Eq. (\ref{eq:state1}) with $\Omega$ given by Eq (\ref{eq:wigangle}).}
\end{figure}

As in Ref. \cite{Terashima}, let us now consider qubits travelling around a circular path that is not necessarily a geodesic orbit. In particular, we consider the qubits moving in the equatorial plane $\theta = \pi/2$ with an angular velocity $\omega$, around the source of the gravitational field. Then, the four-velocity is given by 
\begin{align}
    & u^t = E, \ \ \ u^r = 0, \nonumber \\ 
    & u^{\theta} = 0, \ \ \ u^{\phi} = \omega E,
\end{align}
such that the standard angular velocity is $d \phi/dt = u^\phi / u^t = \omega$. Since $u^{\mu}u_{\mu} = -1$, or equivalently
\begin{equation}
    g_{tt}(u^t)^2 + g_{\phi \phi}(u^\phi)^2 = -f(r) E^2 + r^2 \omega^2 E^2 = -1, \label{eq:norm}
\end{equation}
which implies $E = 1/\sqrt{f(r) - r^2 \omega^2}$. Besides, Eq.(\ref{eq:norm}) suggests a familiar parametrization: $f(r)E^2 = \cosh^2 \xi$ and $r^2 \omega^2 E^2 = \sinh^2 \xi$ such that
\begin{align}
    u^t = \frac{\cosh \xi}{\sqrt{f(r)}}, \ \ \ u^{\phi} = \frac{\sinh \xi}{r}.
\end{align}
Therefore, the non-zero elements of the four-velocity in the local frame defined by the tetrad field are expressed by
\begin{align}
    u^0 = e^0_{\ t}(x)u^t = \cosh \xi, \ \ \ u^{3} = e^{3}_{\ \phi}(x)u^{\phi} = \sinh \xi,
\end{align}
with the speed of the particle in this frame being $v = dx^3/dx^0 = u^3/u^0 = \tanh \xi$, which implies that $r \omega = \sqrt{f(r)} v$ and the familiar expressions $\sinh \xi = v \gamma$ and $\cosh \xi = \gamma$, where $\gamma = (1 - v^2)^{-1/2}$. In order for the particle to maintain such non-geodetic circular orbit, it is necessary to apply an external radial force against gravity and the centrifugal force\footnote{In General Relativity, gravity and the centrifugal force are just manifestations of the spacetime curvature.}, allowing the quanton to travel in the circular orbit with the specific angular velocity $\omega$ at a given distance $r$ from the source. Therefore the non-zero component of the acceleration due to non-gravitational external forces is given by:
\begin{align}
    a^{r}& = u^{\nu} \nabla_{\nu} u^{r} \\
    & = - \frac{\sinh^2 \xi}{r}\Big(1 - \frac{r_s}{2rf(r)} \coth^2 \xi \Big)f(r).
\end{align}
For instance, in the specific case where $u^{\phi} = J/r^2$, which corresponds to a geodetic circular orbit, then $a^r = 0$. The non-zero infinitesimal local Lorentz transformations, defined by Eq. (\ref{eq:infloc}), are
\begin{align}
    & \lambda^{1}_{\ 0}(x) = - \frac{\cosh \xi \sinh^2 \xi}{r}\Big( 1 - \frac{r_s}{2rf(r)}\Big)\sqrt{f(r)}, \\
    & \lambda^{1}_{\ 3}(x) = \frac{\cosh^2 \xi \sinh \xi}{r}\Big( 1 - \frac{r_s}{2rf(r)}\Big)\sqrt{f(r)},
\end{align}
which also corresponds to a boost along the 1-axis and a rotation about the 2-axis. The infinitesimal Wigner rotation associated with the rotation over the 2-axis is given by
\begin{align}
    \theta^{1}_{\ 3}(x) & = \frac{\cosh \xi \sinh \xi }{r}\Big(1 - \frac{r_s}{2rf(r)}\Big) \sqrt{f(r)}\\
    & = \frac{f(r) \omega}{f(r) - r^2\omega^2} \Big(1 - \frac{r_s}{2rf(r)}\Big).
\end{align}
From $\theta^{1}_{\ 3}(x)$, we can calculate the finite local Wigner rotation by integration, such that the finite local Wigner rotation due only to spacetime curvature is given by
\begin{align}
    \Omega := \Theta - \Phi = \Phi \cosh \xi \Big(1 - \frac{r_s}{2rf(r)}\Big) \sqrt{f(r)} - \Phi. \label{eq:wigangle}
\end{align}
It is noteworthy that, as $r \to r_s$, $\Omega$ varies very rapidly such that, at the event horizon, in the strong field limit $\lim_{r \to r_s} \Omega = - \infty$ \cite{Lanzagorta}. In Fig. \ref{fig:e}, we plotted $\Omega$ as function of $r_s/r$ for $\Phi = \pi/8$ and $v/c = 0.1$. As we will see, this fact will cause the complementary aspects of a quanton to oscillate very rapidly near the Schwarzschild radius.

As before, let us consider a pair of entangled spin-$1/2$ particles emitted at given a point on a non-geodesic equatorial circle with the local quantization axis along the 1-axis, as one of the particles of the bipartite state travels in the clockwise direction and the other moves counterclockwise. In other words, we have a pair of entangled particles in opposite directions with constant four-velocity $u^a_{\pm} = (\cosh \xi, 0, 0, \pm \sinh \xi)$ in the state given by Eq. (\ref{eq:state}). After some proper time $\tau = r \Phi/ \sinh \xi$, the particles travelled along their circular paths such that the spinor representation of the finite Wigner rotation due only to gravitation effects is given by $D(W(\pm \Phi)) = e^{\mp \frac{i}{2}\sigma_2 \Omega}$. Therefore, the state of the bipartite system, in the local frame at points $\phi = \pm \Phi$, is also given by Eq. (\ref{eq:state1}). The reduced spin density matrices of each particle are given by
\begin{align}
        & \rho^A_{\Lambda s} = \rho^B_{\Lambda s} = \begin{pmatrix}
\frac{1}{2} & \cos \frac{\Omega}{2} \sin \frac{\Omega}{2}  \ \\
\cos \frac{\Omega}{2} \sin \frac{\Omega}{2} & \frac{1}{2}
\end{pmatrix}, \label{eq:cohe}
\end{align}
and $\rho^A_{\Lambda p} = \rho^B_{\Lambda p} = \frac{1}{2} I_{2 \times 2}$, with $\Omega$ being expressed by Eq. (\ref{eq:wigangle}). In Figs. \ref{fig:f} and \ref{fig:g}, we plotted the behavior $S_l(\rho_{\Lambda s})$ and $C_{hs}(\rho_{\Lambda s})$ as a function of $r_s/r$, for $\Phi = \pi/8$ and $v/c = 0.1$. Hence, for each value of $r \in (r_s, \infty)$, we have a specific value for  $C_{hs}$ and $S_l$.

\begin{figure}[t]
    \centering 
    \subfigure{{\includegraphics[width=8.8cm]{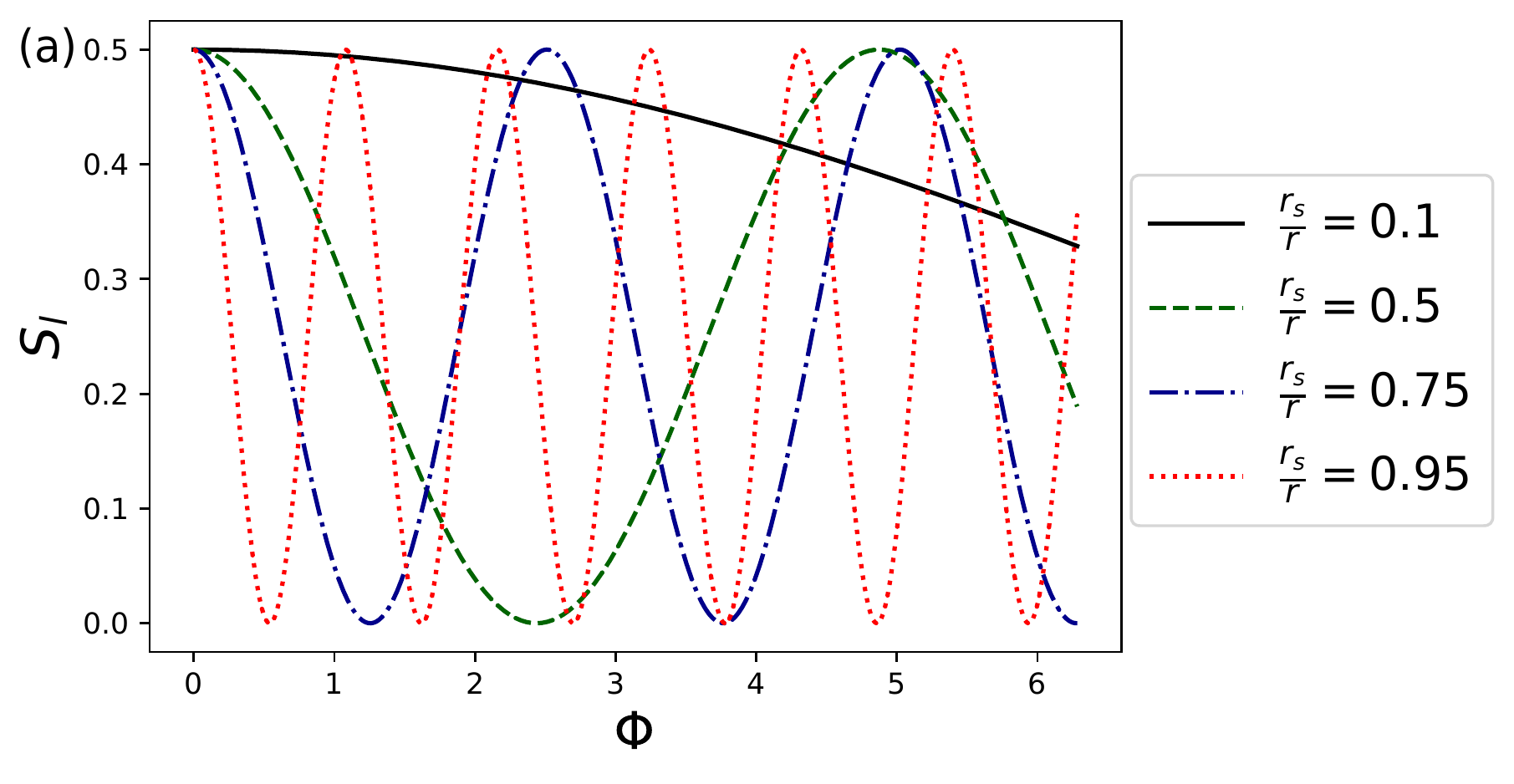}{\label{fig:h}} }}
    \subfigure{{\includegraphics[width=8.8cm]{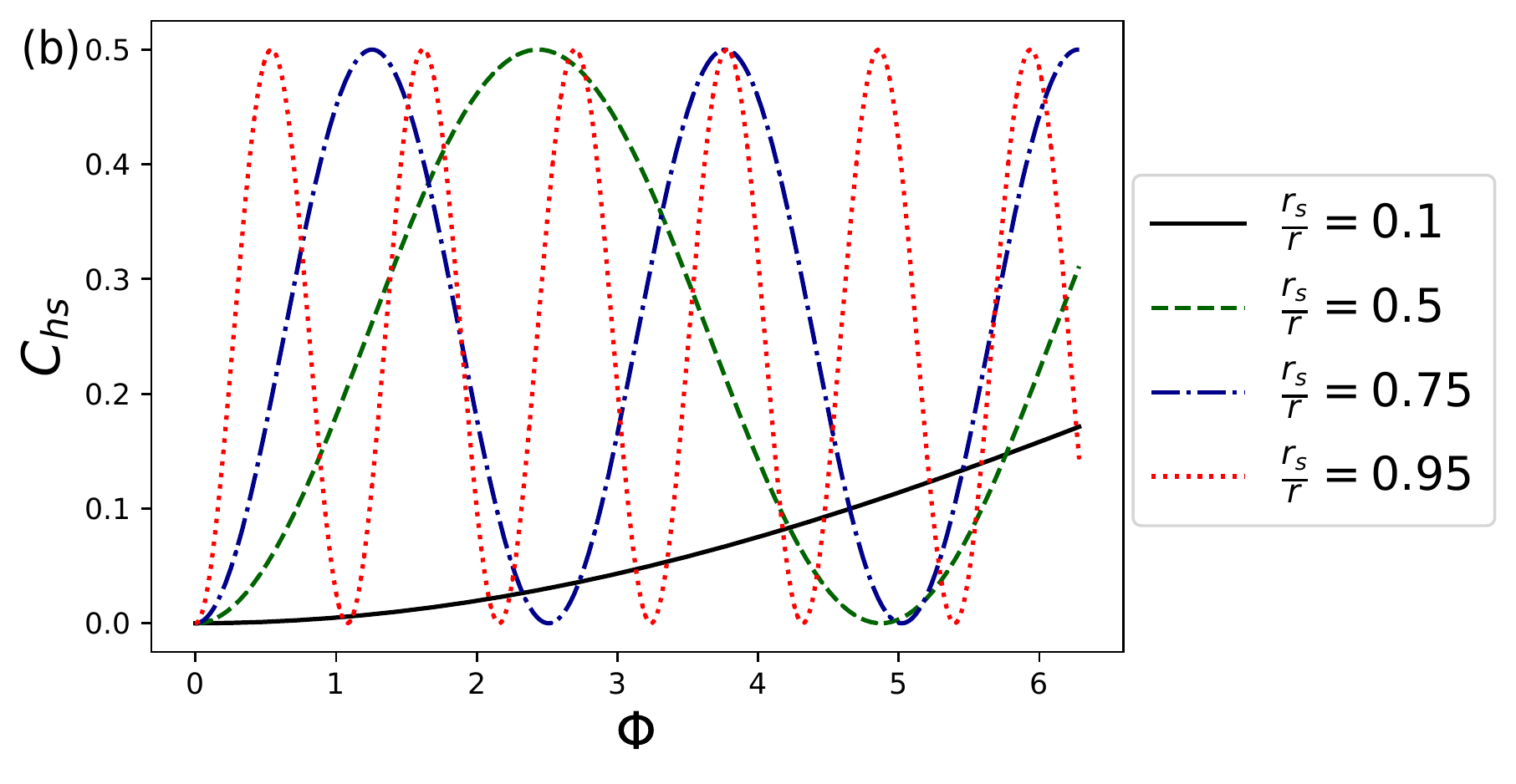}{\label{fig:i}} }}
    \label{fig:example2}
    \caption{(Color online) (a) $S_l$ and (b) $C_{hs}$ as a function of $\Phi$. Quantum coherence and correlation, and their complementarity, for the state in Eq. (\ref{eq:state1}) with $\Omega$ given by Eq (\ref{eq:wigangle}).}
\end{figure}

This rapid oscillation near $r_s$ persists for any value of $\Phi$ and $v$, and it is due to the fact that the Wigner angle varies very rapidly near the Schwarzschild radius. Whereas, in Figs. \ref{fig:h} and \ref{fig:i},  we plotted $S_l(\rho^A_{\Lambda s})$ and $C_{hs}(\rho^A_{\Lambda s})$ as a function of $\Phi$ for different circular orbits. As $\Phi \propto \tau$, these figures express the behavior of $S_l(\rho^A_{\Lambda s})$ and $C_{hs}(\rho^A_{\Lambda s})$ as the particle travels along its circular orbit. One can see that, as $r \to r_s$, the oscillation between a separable and an entangled state of the spin $A$ with the spin of the particle $B$ becomes more and more frequent, which agrees with the fact that the spin precession near $r_s$ is very rapid, as reported in Ref. \cite{Terashima}. This effect is due to the choice of the tetrad field and thus the particular static observer, which becomes singular at the horizon. Therefore, we can conclude that the local static observers (which are not necessarily inertial) attribute a very rapid precession near the horizon. In addition, it is worthwhile mentioning that $S_l(\rho^A_{\Lambda s})$ and $C_{hs}(\rho^A_{\Lambda s})$ can be taken as measures of classical and quantum uncertainties of the spin-state, respectively, since these measures also satisfy the criteria established by Luo \cite{Luo}.

\section{Conclusions}
\label{sec:con}
In this article, we extended complete complementarity relations to curved spacetimes by considering a succession of infinitesimal local Lorentz transformations, which implies that complementarity remains valid as the quanton travels through its world line and the complementarity aspects in different points of spacetime are connected.

This result allowed us to study these different complementary aspects of a quantum system as it travels through spacetime. In particular, we studied the behavior of these different complementary properties of massive spin-$1/2$ particles in the Schwarzschild spacetime. For geodetic circular orbits, we reported that the spin-state of the two particles oscillates between a separable and an entangled state. For non-geodetic circular orbits, we found that the frequency of these oscillations gets bigger as the orbit gets closer to the Schwarzschild radius, which agrees with the fact that the spin precession near $r_s$ is very rapid. This effect is due to the choice of the tetrad field, and thus the particular static observer, as noticed in Ref. \cite{Terashima}. Therefore, the local static observers attribute a very rapid precession of the spin near the horizon. 

On the other hand, if we choose a different vierbein and a different four-velocity (since in this case the orbit is non-geodetic and therefore we can apply a convenient non-gravitational force to choose a specific four-velocity, which differs from the geodetic case, where the four-velocity is specified by the orbit), we can define a different local observer (frame). In particular, it is possible to choose one to avoid the apparent singularity at the horizon. Since this singularity is connected with the breakdown of the coordinate system $(t, r, \theta, \phi)$, it is possible to choose, for instance, the Kruskal-Szekeres coordinate system \cite{Wald}, in which the metric is not singular at $r_s$. Then, it is possible to construct a different vierbein such that the local Wigner rotation does not diverge at $r = r_s$. This was done by Terashima and Ueda in Ref. \cite{Terashima}, where they studied EPR-correlations near the horizon. However, in the new tetrad-field related to the Kruskal-Szekeres coordinates, the new local frame falls into the black hole. Therefore, to perform measurements in this local frame, the observers also must fall into the black hole. As well, in this case, the authors used a four-velocity such that the particle also falls in the black hole. Finally, the authors concluded that this precession is not singular at the horizon, and the observers on the horizon can extract the EPR-correlation. 

In contrast, our work goes into another direction. We just wanted to show that complete complementarity relations are valid in curved spacetimes for different observers, and explore some examples, which corresponds to static observers. For instance, for circular geodesics this is not a problem, since stable circular orbits are those in which $r>1.5r_s$. But, as well, for circular orbits close to $\frac{3}{2}r_s$, the precession is larger than for those orbits far away of $\frac{3}{2} r_s$. Besides, it is worth mentioning that this dependence on the tetrad-field and consequently on the local observer is not so surprising, since, it is well known that for an observer in radial free-fall in Schwarzschild spacetime, the observer will take a finite interval of proper time to pass the event horizon and reach the singularity at $r = 0$. Whereas, for the coordinate frame in the infinity, defined by the coordinates $(t, r, \theta, \phi)$, the free-fall observer will take an infinite amount of time \cite{Carroll}.

In addition, we discussed the fact that the effect of spacetime curvature in the complementary behavior of these quantum states is analogous to the effect reported in Ref. \cite{Zych}, since the clockwise and counterclockwise circular paths can be taken as the different paths of a Mach-Zehnder interferometer. Hence, our work helps in the understanding of how the spacetime curvature affects the behavior of these complementary properties of a quantum system, as well it opens the possibility for different studies. For instance, it is possible to explore different spacetime geometries, and how these different geometries affect the complementary behavior of a quanton. Also, we did not took into account the spin-curvature coupling, which need to be taken into account when investigating this relation in the case of supermassive compact objects and/or ultra-relativistic test particles.

\begin{acknowledgments}
This work was supported by the Coordena\c{c}\~ao de Aperfei\c{c}oamento de Pessoal de N\'ivel Superior (CAPES), process 88882.427924/2019-01, and by the Instituto Nacional de Ci\^encia e Tecnologia de Informa\c{c}\~ao Qu\^antica (INCT-IQ), process 465469/2014-0.
\end{acknowledgments}


\end{document}